\documentclass[%
 reprint,
 amsmath,amssymb,
 aps,
]{revtex4-1}

\usepackage{graphicx}
\usepackage{dcolumn}
\usepackage{bm}

\newcommand{\ket}[1]{|#1\rangle}
\newcommand{\expt}[1]{\langle#1\rangle}

\begin{document}


\title{A Multiple Observer Probability Analysis for Bell Scenarios in Special Relativity}



\author{ Kevin Vanslette}
%

\date{\today}

\begin{abstract}
Here we present a Multiple Observer Probability Analysis (MOPA) for the purpose of clarifying topics in experimental Bell scenarios. Because Bell scenarios are interested in quantum effects between nonlocal measurement devices, we assign an observer to each device: Alice and Bob. Given that the observers are stationary and space-like separated, each observer is privy to different information along their shared equi-temporal lines due to permutations in the order they observe events. Therefore, each observer is inclined to assign different probability distributions to the same set of propositions due to these informational differences. The observers are obligated to update their probability distributions on the basis of locally observed events, and in this sense, factuality is informational locality. In this framework, only local variables or detections may be factual, but nothing prevents an observer from inquiring or making if-then inferences on the counterfactual basis of a nonlocal proposition being true. Indeed the objects pertaining to these nonlocal counterfactual propositions may be far outside an observer's light cone. The MOPA arrives at the conclusion that the CHSH inequality is only nonlocally violated \emph{counterfactually} by each observer whereas local violations of the CHSH may be factual or counterfactual. We believe the MOPA to better gel probability theory (and thus QM) with Special Relativity than does the standard locality conditions imposed in the Bell and CHSH inequalities. The \emph{no-signaling} condition is reinterpreted, and perhaps further clarified, in the MOPA and statements about counterfactuality and observer dependent QM are made. 
\end{abstract}

\maketitle


\section{\label{sec:level1}Introduction}
The primary realization of Einstein in Special Relativity (SR) was that the notion of simultaneity depends on the relative motion of an observer with respect to space-time events. The lack of a universal notion of simultaneity ultimately leads to the observed effects of length contraction and time dilation.   Although observers can co-ordinate and ultimately agree on the location of events and intervals in space-time, each observer rides their own world line and observes events from their own perspective. Because the \emph{observed order} of space-like separated events and observed timing of events depend on an observers relative motion and location in space-time, different world lines and thus \emph{different observers are privy to different information at their respective space-time locations}.

This naturally leads to the notion of observer dependent probability distributions; observer $A$ assigns probability distributions such as,
\begin{eqnarray}
P_A(a,b,c,...|I_A),
\end{eqnarray}
to the propositions $a,b,c,...$ given their knowledge of previously observed and/or known information $I_A$. Observer $B$, by following a different world line or having a different informational background than $A$, has the information $I_B$, and assigns a potentially different probability distribution,
\begin{eqnarray}
P_B(a,b,c,...|I_B),
\end{eqnarray}
to the same set of propositions. Observer dependent probability analysis of this type will be called a Multiple Observer Probability Analysis (MOPA).

The natural and evolutionary solution to informational inequivalence between observers is communication. Full communication between observers leads to a pooling of known information. This may be represented by $I_{A\cup B}=I_A\cup I_B$, and thereby the observer dependent probability distributions $P_A(a,b,c,...|I_A)\rightarrow P_{A\cup B}(a,b,c,...|I_{A\cup B})$ and  $P_B(a,b,c,...|I_A)\rightarrow P_{A\cup B}(a,b,c,...|I_{A\cup B})$ equilibrate. A realist interpretation of the universe requires two observers to agree on the outcome of events, such that the propositions of $I_A$ and $I_B$ cannot conflict; therefore, the difference in information between $I_A$ and $I_B$ cannot lead to analysis such as $I_{A\cap B}=a\wedge \tilde{a}$ -- the illogical statement that $a$ and its negation, $\tilde{a}$, are both true. Because observers in general do not agree on the lengths, times, and order of events due to SR, part of the propositions $I_O$ known to observer $O$, must include, in conjunction, the knowledge of his or her's own world line with reasonable certainty such that Lorentz (and other) transformations may be applied to avoid anti-realism -- this is aptly stated in \cite{book} as ``a system of coordinates carry no information". This allows observer $A$, for instance, to observe one frequency of light while another observer $B$, in relative motion to $A$, observers a different frequency of light from the same source without breaking realism. The proper account of the information known to an observer requires a great deal of attention to detail, as is displayed in the difficulties underlying the Monte-Hall, and like, probability problems. We take inspiration from Einstein and assume the laws of probability and probability updating are ubiquitous among observers, while an observer's particular assigned probability distribution may differ.

 The struggle with publishing the correct application of probability theory to problems in Physics is ultimately that they lead to the same correct conclusions, and therefore nothing ``new" is learned. In this sense, this article is a review of Bayesian probability theory applied to Bell type scenarios in quantum physics.  What is interesting is that the ``incorrect" use of Bayesian probability theory, which ultimately leads to Bell inequalities, has generated a vast literature in telling us precisely how \emph{not} to apply probability theory rather than the statement of how probability theory \emph{should} be applied. We will review some of the arguments that lead to Bell-like inequalities, and then using a MOPA, show the sense in which Bell's inequality is \emph{not} derived. This is certainty an odd type of result, but in completing this exercise, we have found a natural setting for Bayesian probability theory and data analysis in SR -- which includes the probability analysis of QM as a special case. 

 As two space-like separated observers (Alice and Bob) with respective measurement devices are needed to complete a Bell scenario type experiment, the MOPA offers a natural experimental account of both observers, who, due to the retarded propagation of signals, disagree on the observed order of events. 
In this framework, only local variables or detections may be factual, but nothing prevents an observer from inquiring or making if-then inferences on the counterfactual basis of a nonlocal proposition being true. Using a MOPA, we find that nonlocal violations of the CHSH inequality only occur counterfactually as Alice (Bob) is not privy to the measurement setting of Bob (Alice). Local violations of the CHSH inequality using a MOPA come from either Alice and Bob agreeing to preset measurement settings before the experiment or communicating measurement settings afterward. We find a number of observer dependent probability distributions that may be useful for inference, communication, and quantum steering. 

We expect the MOPA to be, more or less, notionally compatible with interpretations of QM that are probabilistic in nature, e.g. Entropic Dynamics \cite{EDNew} and QBism \cite{QBism}, while also being compatible with interpretations that already account for multiple observers and event order within their framework \cite{PL}. The MOPA is also operationally compatible with all other interpretations of QM that admit collapse and probability updating.
\subsection{A Note on Information, Bayes Theorem, and Marginalization:}
 In instances when the truthiness of a proposition, $b$, is learned and happens to be correlated to $a$, one is obligated, for the purpose of making informed judgments, to update an old probability $p(a)$ to a new probability,
\begin{eqnarray}
p(a)\stackrel{*}{\rightarrow} p(a|b)\label{update}.
\end{eqnarray}
From this point of view it is natural to define \emph{information} operationally $(*)$ as the \emph{rationale} that causes a probability distribution to change (inspired by \cite{book}), (\ref{update}) being one example. Given $b$ is true, the probability of $a$ conditional on $b$, $p(a|b)$, may be rewritten using the product rule (where logically $p(b)>0$),
\begin{eqnarray}
p(a|b)=\frac{p(b|a)p(a)}{p(b)},\label{BayesTheorem}
\end{eqnarray}
which is Bayes Theorem. If it is unclear whether $b$ is true or its logical negation, $\tilde{b}$, is true, then one can use the fact that $b$ or its negation, $b\vee\tilde{b}$, is certainly true, and make the following inference about $a$,
\begin{eqnarray}
p(a,b\vee\tilde{b}|d)=p(a,b|d)+p(a,\tilde{b}|d)=\sum_Bp(a,b|d)\nonumber\\
=p(a|d)\sum_{B}p(b|a,d)=p(a|d).
\end{eqnarray}
This process is called \emph{marginalization}, and in particular above, $b$ has been marginalized over to give a best guess for $a$. Marginalized probabilities, such as $p(a|d)$, may equally be read as $p(a,b\vee\tilde{b}|d)$, the probability of $a$ and ($b$ or not $b$). Furthermore, using the product rule, one finds that a marginalized distribution may be further restated as,
\begin{eqnarray}
p(a,b\vee\tilde{b}|d)=p(a|b\vee\tilde{b},d)p(b\vee\tilde{b}|d)=p(a|b\vee\tilde{b},d),
\end{eqnarray}
meaning $p(a|d)=p(a|b\vee\tilde{b},d)$, which is read the probability of $a$ given ($b$ or not $b$). 

\section{\label{sec:level2} CHSH, No-Signaling, and MOPA}
A probabilistic approach to Bell's Theorem \cite{BellEPR} was given by Clauser, Horne, Shiminy, Holt (CHSH) \cite{CHSH}, and an excellent review of the current state of affairs is provided by \cite{CavalcantiBellReview} and the references therein. Bell's Theorem has manifested itself over the years in several forms: a Bayesian account is given by \cite{Garrett}, proofs of contextuality and the Bell-Kochen-Specker Theorem \cite{Bell,KS,Mermin, Ballentine}, and more recently the inconsistency of Causal Models with QM \cite{SpekkensCausality,Cavalcanti}. In the CHSH representation of Bell's inequality, one represents a local hidden variable theory (or a causal account) of a Bell scenario with a factorizability condition,
\begin{eqnarray}
p(a,b|x,y,\lambda)\stackrel{fac}{=}p(a|x,\lambda)p(b|y,\lambda),\label{factor}
\end{eqnarray}
such that the outcome $a$ ($b$) is independent of the nonlocal measurement setting $y$ ($x$) that measures $b$ ($a$), where both outcomes may be coupled through a hidden variable, or common cause, $\lambda$. As is pointed out in \cite{SpekkensCausality}, the common cause $\lambda$ could itself be the wavefunction and the CHSH is still violated. Directly from \cite{CavalcantiBellReview}, ``This factorizability condition simply expresses that we
have found an explanation according to which the probability
for $a$ only depends on the past variables $\lambda$ and on
the local measurement $x$, but not on the distant measurement
and outcome, and analogously for the probability
to obtain $b$.".  Marginalizing over the hidden variable of a joint probability distribution compounded out of the product of (\ref{factor}) and $p(\lambda)$ is,
\begin{eqnarray}
p(a,b|x,y)=\int_{\Lambda}d\lambda \,p(\lambda)p(a|x,\lambda)p(b|y,\lambda).\label{margfactor}
\end{eqnarray}
Using the above probability distribution, one can express the CHSH inequality,
\begin{eqnarray}
S_{\lambda}\equiv\expt{ab}_{00}+\expt{ab}_{10}+\expt{ab}_{01}-\expt{ab}_{11}\leq 2,\label{CHSH}
\end{eqnarray}
where $\expt{ab}_{xy}=\sum_{ab} ab\,p(a,b|x,y)$ takes values between $[-1,1]$, for measurement settings $x,y=\{0,1\}$. Paradoxical to our intuition, QM is able to violate the CHSH $S_{QM}=2\sqrt{2}$, while the PR-box $S_{PR}=4$ maximally violates the inequality \cite{CHSH,CavalcantiBellReview}. It should be noted that it is a feature of the generality of probability theory that allows us to formulate probabilities, such as in the PR-box, that may or may not bear physical relevance or be properly constrained by physics.

Bell scenarios in QM and the PR-box obey the so called \emph{no-signaling} condition \cite{Cirelson,CavalcantiBellReview},
\begin{eqnarray}
\sum_bp(a,b|x,y)\equiv p(a|x,y)\stackrel{n.s.}{=}p(a|x),\nonumber\\
\sum_ap(a,b|x,y)\equiv p(b|x,y)\stackrel{n.s.}{=}p(b|y),\label{nosignal}
\end{eqnarray}
for all $a,b,x,y$, but, as is evident from the CHSH violations, the probabilities do not obey the factorization condition (\ref{factor}). Strictly speaking, from a probability point of view, the \emph{no-signaling} condition, in contradiction with its name, \emph{does not in general represent no-signaling}.

What is actually represented in (\ref{nosignal}) is something like \emph{marginal measurement setting independence}, that is, seemingly \emph{knowing} the measurement setting $y$ does not change our state of knowledge $p(a|x,y)\rightarrow p(a|x)$ about $a$. This is evident from the fact that the measurement setting $y$ appears in the conditional part of the probability $p(a|x,y)$, and in this sense, \emph{it has been signaled} as its value is given, unless otherwise clarified (as we will in the MOPA). Furthermore, marginalizations over detection settings can be confused with the ``\emph{no-signaling}" condition because notationally,
\begin{eqnarray}
p(a|x,y)\stackrel{n.s.}{=}p(a|x)\nonumber
\end{eqnarray}
\begin{eqnarray}
\sim \sum_yp(a,y|x)=\sum_yp(a|x,y)p(y)=p(a|y\vee\tilde{y},x)=p(a|x),\nonumber\\
\end{eqnarray}
the two lines appear to be equivalent when in fact they are only equal iff $p(a|x,y)=p(a|x)$ is independent of $y$, for all $y$, which is not true in general; however, it is true for Bell scenarios in QM and the PR-Box. Equation (\ref{nosignal}) is still useful for determining, not \emph{no-signaling} but rather \emph{marginal measurement setting independence}, which is interesting in QM in its own right. Due to the possibility of misrepresenting \emph{no-signaling} through (\ref{nosignal}), we represent \emph{no-signaling} in the MOPA simply in terms of no faster than light signaling.  


As there are multiple observers in a MOPA, a natural inquiry about (\ref{factor})-(\ref{nosignal}) is ``who has assigned these distributions... Alice, Bob, Charlie, or is each observer contributing a piece?" -- the latter appears to be the case in (\ref{factor}), but it is difficult to know at this point, and it should be noted that probability distributions from different observers are not normally combined via a product rule. 

Quantum Mechanics generates probability distributions that seemingly imply observers have access to the value of nonlocal measurement settings at all times.  When considerations of relativity are finally made, it is sometimes concluded that quantum mechanics must be nonlocal rather than the reverse, that the analysis has failed to properly capture an observer's local information.  In reality, the assumptions of quantum theory and relativity are at odds as much as Newtonian mechanics and Maxwell's theory were at odds before the notion of simultaneity was understood -- we hope to further clarify probability theory in SR here.

\section{\label{sec:level3} MOPA of Bell Scenarios}
As is motivated in the introduction, observer dependent information exists. In the Bell scenarios, an observer's locally available information is their measurement settings and measurement outcomes (Alice may know hers and Bob his, but before communication, neither know each-others). We give detailed experimental accounts of Bell scenarios in the MOPA by Alice and Bob and track how the statistics of the experiment change throughout the measurement process. The analysis makes it apparent who knows and who cannot know what-when and shows that the seemingly natural factorization condition is less obvious than might be expected due the asymmetrical order of observations made by Alice and Bob. 

It is notationally convenient to denote a quantity that has been locally measured, or a proposition that is known factually to be true, to follow two vertical bars in a probability distribution, while placing counterfactual ``if true" propositions between two vertical bars such that,
\begin{eqnarray}
p(a|b,y)\rightarrow p(a|b|y),
\end{eqnarray}
is read, ``the probability of $a$ \emph{if} $b$ takes the value $b$ and given $y$ was measured (or is known) to be $y$ (locally)". For calculation purposes the second bar may be ignored (or replaced by a comma) as $p(a|b|y)=p(a|b,y)$ in value but not interpretation. This helps communicate whether a proposition in a probability distribution is counterfactual ($b$) or factual ($y$). In this framework, only local variables or detections may be factual and follow two bars, but nothing prevents an observer from inquiring or making if-then inferences on the counterfactual basis of a nonlocal proposition being true.

The generality of probability theory lets observers inquire about the probability of things that may or may not happen. For instance, \emph{if} in the future I get a large pay increase $+$ at my job $J$ then I will probability buy a sports car $c$, that is $p(c|+|J)\sim 1$, when in-fact no pay increase has been given, I do not own a sports car, but lucky I do, in-fact, still have a job.

The MOPA, and the divvying-up of propositions into the factual and counterfactual, leads to the assessment that nonlocal violations of the CHSH inequality can only occur counterfactually, and in this sense, \emph{counterfactuality is itself informational nonlocality} -- one is able to consider counterfactual changes in a probability distribution based on nonlocal counterfactual propositions. Indeed the objects pertaining to these nonlocal counterfactual propositions may be well outside an observer's light cone. 
\subsection{Initial conditions: $t_0$}
In Bell scenarios there are four main points in time $t_0<t_{\theta}<t_{\pm}<t_c$: where $t_0$ the initial time, $t_{\theta}$ is the time the measurement setting $\theta$ is chosen by their respective observer, $t_{\pm}$ the time a measurement outcome  $\pm$ is made, and $t_c$ is the latest time when the results of measurement outcomes and measurement settings are communicated for statistical analysis. In general each time coordinate $t$ should have an observer label, but because the observers are stationary in the lab frame, this index is suppressed. It is assumed that at the time $t_0$, both space-like separated stationary observers Alice and Bob know with certainty the state of the system $\psi_0$, where $\psi_0$ may in general represent a classical, quantum, or PR-box type system. At $t_0$, we let Alice and Bob have equal information about the state of the system, and thus they assign equivalent joint probability distributions,
\begin{eqnarray}
P_A(\pm_{a},\theta_a,\pm_{b},\theta_b||\psi_0,t_0)=P_B(\pm_{a},\theta_a,\pm_{b},\theta_b||\psi_0,t_0),\nonumber\\\label{t_0}
\end{eqnarray}
to the set of propositions $\{\pm_{a},\theta_a,\pm_{b},\theta_b\}$. At this time there are something like 60 representations of the same joint probability distribution by the product rule in all possible fashions (permute propositions about the first vertical bar and apply the product rule and permutations recursively). Because nothing has been measured by either observer, these representations are in terms of the products of counterfactually conditional probability distributions. Each unique counterfactual conditional probability distribution (as well as each marginalization) represents a question that may asked by the observer without breaking locality. Locality would be broken if a nonlocal outcome or measurement setting of the space-like separated event was assumed to take definite factual values at $t_0$.

Of these possible counterfactual probabilistic inquiries, perhaps the most relevant at $t_0$ is,
\begin{eqnarray}
P_A(\pm_{a},\pm_{b}|\theta_a,\theta_b|\psi_0,t_0)=P_B(\pm_{a},\pm_{b}|\theta_a,\theta_b|\psi_0,t_0).\nonumber\\\label{14}
\end{eqnarray}
These distributions presumably take the same value, and in the case of an anti-symmetric Bell state,
\begin{eqnarray}
P_{QM}(\pm_{a},\pm_{b}|\theta_a,\theta_b)=\frac{1}{4}+\frac{\delta_{\pm_a,\mp_b}-\delta_{\pm_a,\pm_b}}{4}\cos(\theta_a-\theta_b),\nonumber\\
\label{QMBell}
\end{eqnarray}
(where notationally $\mp_b=-\pm_b$ and $\theta_a,\theta_b$ are the angles of Alice and Bob's measurement devices, respectively), the observer dependent probability distributions may adopt the functional form of $P_{QM}$ and therefore violate the CHSH inequality. However, because at $t_0$ Alice and Bob do not know each other's and haven't chosen their own measurement settings, at best the values of the measurement settings can be posited counterfactually. Therefore (\ref{14}) reads, ``Alice's (Bob's) probability for $\pm_a$ and $\pm_b$ at $t_0$ given the system is factually $\psi_0$ and \emph{if} the measurement settings are set to $\theta_a$ and $\theta_b$, has the probability...".  Thus at $t_0$, Alice and Bob's probability distributions will infact violate the CHSH inequality, \emph{but only counterfactually} since the actual values of $\theta_a$ and $\theta_b$ are not known and are simply posited. It is possible for Alice and Bob to posit these counterfactual probabilistic inquiries because there is no physical issue in inquiring about nonlocal events or possible \emph{if-then} scenarios -- the issue is that without proper specification, counterfactually known propositions can be mistaken for factually known propositions. At the time $t_0$ it is now clear that the CHSH inequality is only violated counterfactually. 

The only way that $\theta_a$ and $\theta_b$ could be known factually by Alice and Bob is if Alice and Bob had communicated ahead of time and preset their measurement settings. This situation does not capture the spirit of Bell's Theorem and the CHSH, as the violation would be local (and therefore not exude ``quantum nonlocal weirdness") as Alice and Bob had communicated ahead of time.

At this time we should really reevaluate the CHSH inequality. Because the CHSH inequality is a sum of expectation values over different measurement settings, it should be noted that a single experiment can in no way factually violate the CHSH as any experiment only has one factual set of measurement settings. By ``nonlocally violating the CHSH counterfactually at $t_{0}$" we simply mean that the functional form of $P_A(\pm_{a},\pm_{b}|\theta_b\theta_a|\psi_0,t_{0})$ is the same as (\ref{QMBell}), but with $\theta_a,\theta_b$ forcibly being counterfactual by SR. To properly conform to probability theory, the CHSH inequality is better stated as an expectation value itself,
\begin{eqnarray}
\sum_{\pm_a,\theta_a,\pm_b,\theta_b}abP_A(\pm_{a},\theta_a,\pm_{b},\theta_b||\psi_0,t_0)\nonumber
\end{eqnarray}
\begin{eqnarray}
=\frac{1}{4}\Big(\expt{ab}_{00}+\expt{ab}_{10}+\expt{ab}_{01}-\expt{ab}_{11}\Big)=\frac{1}{4}S_{\lambda}\leq \frac{1}{2},\nonumber\\\label{CHSHexpt}
\end{eqnarray}
and therefore $S_{\lambda}\leq2$ for $P_A(\theta_a,\theta_b)=\frac{1}{4}$ over the measurement settings $\theta_a,\theta_b$ in $\{0,1\}$. This immediately shows that the CHSH inequality is not something that can be factually verified in a single experiment as multiple measurement settings are required. Strictly speaking, the CHSH inequality may only be violated for a single experiment before the measurement settings are known factually ($P_A(\theta_a,\theta_b)<1$ for all $\theta_a,\theta_b$), and therefore nonlocal violations of the CHSH inequality are completely counterfactual as the angle dependencies in the probabilities are counterfactual.

\subsection{Alice chooses her measurement setting: $t_{\theta}$ \label{sect_0}}
Alice, being privy to the settings of her measurement device, sets the angle to be $\theta_a$ mid flight $t_0<t_{\theta}<t_{\pm}$. Her state of knowledge before detection is therefore described by the joint probability distribution,
\begin{eqnarray}
P_A(\pm_{a},\pm_{b},\theta_b||\theta_a,\psi_0,t_{\theta}).\label{8}
\end{eqnarray}
Again, the counterfactual probabilistic inquiry of interest is
\begin{eqnarray}
P_A(\pm_{a},\pm_{b}|\theta_b|\theta_a,\psi_0,t_{\theta}),\label{8}
\end{eqnarray}
which is Alice's probability of $\pm_{a}$ and $\pm_{b}$ given her setting is indeed at $\theta_a$ \emph{if} Bob's measurement device takes the value $\theta_b$. The distributions at $t_{\theta}$ again is numerically equal to (\ref{QMBell}), but the values of $\theta_b$ are at best be posited counterfactually. Again, the only way that $\theta_b$ could be known factually by Alice is if Alice and Bob had communicated its value ahead of time, which again, would not capture the spirit of the CHSH.

If Alice has measurement uncertainty in her angle $\theta_a$, the uncertainty may be quantified by the probability distribution $Q(\theta_a)$, which perhaps is a narrow Gaussian about $\theta_a$. Given that is the case, Alice may write the following joint probability distribution at $t_{\theta}$,
\begin{eqnarray}
P_A(\pm_{a},\pm_{b},\theta_a,\theta_b||\psi_0,t_{\theta})\nonumber\\
=P_A(\pm_{a},\pm_{b},\theta_b||\theta_a,\psi_0,t_{\theta})Q(\theta_a),\label{PQtheta}
\end{eqnarray}
which is (\ref{8}) multiplied by the measurement uncertainty $Q(\theta_a)$. In all cases of interest, local quantities that may in principle be known ``factually" follow two vertical bars and may be multiplied by a measurement uncertainty distribution $Q$ to form the full joint probability distribution after a detection has been done. If the measurement is precise, then $Q$ may be a Kronecker or Dirac delta function at the detected value. It should be noted that Alice is in no way forced to marginalize over Bob's device setting or measurement outcomes -- she may marginalize over them if she chooses to make further inquiries on that basis.

In the peculiar situation in which Alice knows she has set the measurement angle to $\theta_a\in\{0,1\}$ but has failed to verify which value $\theta_a$ actually is, then again (\ref{PQtheta}) may be used to counterfactually violate (\ref{CHSHexpt}). In this case Alice is in the unusual predicament where she knows the measurement setting has a definite value within her local light cone but the precise information is unaccessible for whatever reason (perhaps she heard a click but that did not specify $\theta_a$). We will continue to denote the value of $\theta_a$ behind double bars as it is ``local in her light cone", but in principle it does not hurt to also consider its precise value as counterfactual as, in some sense, Alice is not within the ``signal velocity-time" cone (or paths) of the measurement setting's signal.

\subsection{Alice makes her measurement: $t_{\pm}$}
If Alice detects the spin value of an entangled particle measured along $\theta_a$, then there are less counterfactual inquires she can make. Her updated distribution of the entire system is therefore,
\begin{eqnarray}
P_A(\pm_{b},\theta_b||\pm_{a},\theta_a,\psi_0,t_{\pm}),
\end{eqnarray}
and there are only 2 counterfactual questions that may be asked at this time,
\begin{eqnarray}
P_A(\pm_{b}|\theta_b|\pm_{a},\theta_a,\psi_0,t_{\pm}),\label{b|y}\\
P_A(\theta_b|\pm_{b}|\pm_{a},\theta_a,\psi_0,t_{\pm}),\label{y|b}
\end{eqnarray}
and 2 marginals,
\begin{eqnarray}
P_A(\pm_{b}||\pm_{a},\theta_a,\psi_0,t_{\pm}),\\
P_A(\theta_b||\pm_{a},\theta_a,\psi_0,t_{\pm}).
\end{eqnarray}
Using the counterfactual probability distributions (\ref{b|y}) and (\ref{y|b}) Alice can guess what Bob's device will read out \emph{if} Bob chooses the angle $\theta_b$ or guess his most probable angle \emph{if} Bob detects $\pm_b$. These are specific examples of Alice using Bayes Theorem to make inferences about what Bob will get based on her local measurements.

Usually Alice can measure $\pm_a$ more or less with complete certainty $Q(\pm_a)=\delta_{\pm_a,\pm_a'}$, but she will have some uncertainty $Q(\theta_a)$ about the measurement setting. Her new joint probability distribution over the whole system at $t_{\pm}$ is,
\begin{eqnarray}
P_A(\pm_{a},\theta_a,\pm_{b},\theta_b||\psi_0,t_{\pm})\nonumber\\
=P_A(\pm_{b},\theta_b||\pm_{a},\theta_a,\psi_0,t_{\pm})Q(\theta_a)Q(\pm_a).
\end{eqnarray}
At this time it is noteworthy to consider the counterfactual probabilistic inquiry,
\begin{eqnarray}
P_A(\pm_{a},\pm_{b}|\theta_b|\theta_a,\psi_0,t_{\pm})\nonumber\\
=Q(\pm_a)P_A(\pm_{b}|\theta_b|\pm_{a},\theta_a,\psi_0,t_{\pm}).
\end{eqnarray}
This distribution only obeys a subset of the statistics that is obeyed by (\ref{QMBell}), simply because Alice has actually measured the outcome of her measurement device to be $\pm_a'$. This is not an issue with the analysis, rather the process is overly describing the whole measurement and detection process, which therefore involves ``collapse" when a detection is made, changing the statistics. 

Again Alice could only nonlocally violate the CHSH counterfactually. If one wishes, similar arguments to last paragraph in Section B could be made about Alice not knowing exact values of $\pm_a,\theta_a$ to formulate the nonlocal violation of the CHSH counterfactually for this single trial.  
\subsection{Bob communicates his measurement angle, his measurement outcome, or both to Alice: $t_c$}
At this latest time $t_c$, the process of communication completely determines the outcomes of the experiment so in principle everything is known: Bob tells Alice $\theta_b$ (with uncertainty $Q_B(\theta_b)=Q_A(\theta_b)$ given the communication channel is noiseless), as well as for $\pm_b$. Alice and Bob agree on the final joint probability distribution after communication,
\begin{eqnarray}
P_A(\pm_{a},\theta_a,\pm_{b},\theta_b||\psi_0,t_c)\nonumber\\
=P_B(\pm_{a},\theta_a,\pm_{b},\theta_b||\psi_0,t_c)\nonumber\\
=Q(\theta_a)Q(\pm_a)Q(\theta_b)Q(\pm_b).\label{data}
\end{eqnarray}
which represents the state of knowledge of a single outcome of the experiment, as their information is pooled $I_{A\cup B}$.   Because the $Q$ distributions represent macroscopic measurement uncertainty (not weak measurement) they are well behaved in the sense that the experimental \emph{data} is the unique set of propositions which maximize (\ref{data}), denoted $d_i=\{\theta_a',\pm_a',\theta_b',\pm_b'\}$ -- the data for experiment $i$. Repeating this experiment $N$ times allows one to estimate the probability distribution (\ref{t_0}) via its frequency using a multinomial distribution. The inferred distribution may be used to estimate the statistical expectation values of (\ref{t_0}) with uncertainty $\propto \frac{1}{\sqrt{N}}$ by the central limit theorem. Tomographically complete data allows for the sampling (or experimental) distribution to span the full probability space of the statistical distribution -- meaning all values of $\theta_a,\theta_b$ should be inspected to fully describe the system. It should be noted that because the state has effectively collapsed, the $n$-moment set of expectation values of (\ref{data}) in no way match the expectation values from (\ref{t_0}). Communication between Alice and Bob is required for experimental verification, and therefore the full set of information becomes local for both observers. 

Learning Bob's measurement angle and outcome, Alice may retrodictively explain the likelihood of her measurement outcome. In some instances, these kind of two-time inferences may be written using a factual current time and a secondary counterfactual time -- Alice's retrodictive inference of the likelihood of her measurement outcome after $t_{\theta}$ may be stated as $P_A(\pm_a|t_{\theta}|\theta_a,\pm_b,\theta_b,t_c)$, as the information is known factually at $t_c$.

The probability analysis for Bob is the same as Alice, except with $a$'s and $b$'s swapped due to the simple symmetry in the problem. The primary probability distributions of interest are listed in the table below:
\begin{center}
\begin{tabular}{c|c|c|c| }

   & Alice & Bob & =?\\ 
\hline
 $t_0$ & $P_A(\pm_{a},\theta_a,\pm_{b},\theta_b||\psi_0,t_0)$ & $P_B(\pm_{a},\theta_a,\pm_{b},\theta_b||\psi_0,t_0)$ & y \\  
\hline
$ t_{\theta}$ & $P_A(\pm_{a},\pm_{b},\theta_b||\theta_a,\psi_0,t_{\theta_a})$ & $P_B(\pm_{a},\theta_a,\pm_{b}||\theta_b,\psi_0,t_{\theta_b})$ & n\\  
\hline
$t_{\pm}$ & $P_A(\pm_{b},\theta_b||\pm_{a},\theta_a,\psi_0,t_{\pm})$ & $P_B(\pm_{a},\theta_a||\pm_{b},\theta_b,\psi_0,t_{\pm})$ & n \\  
\hline
$t_c$ & $Q(\theta_a)Q(\pm_a)Q(\theta_b)Q(\pm_b)$ & $Q(\theta_a)Q(\pm_a)Q(\theta_b)Q(\pm_b)$ & y \\  
\hline
\end{tabular}
,
\end{center}
The last column assigns a yes, $y$, or a no $n$ if the distributions are equal or not equal.  Counterfactual conditional and marginal distributions may be generated from these joint distributions at their respective times and measurement uncertainty may be included on factual propositions. Other experimental designs, which might involve reversing or omitting observations at $t_{\theta}$ or $t_{\pm}$ by either observer are also possible.
\subsection{Discussion}
Provided above is a MOPA for Alice and Bob in a Bell scenario, where the lack of a consistent observed order of events, from SR, has been assumed and accounted for probabilistically. Locality in the MOPA is simply represented by the particular order in which Alice or Bob are signaled information, and when received, update their distributions factually. Because the full factual account of an experiment requires all of the information to be local (i.e. communicated), there is no experiment that allows for the testing of nonlocal signaling, as any nonlocal manipulations cannot be observed, by the fact that all signals are local. As has been highlighted through this analysis, this does not prevent nonlocal observers from making counterfactual inferences or inquiries, and it is in this sense that locality is only violated \emph{counterfactually}. In this analysis QM has been assumed, and yet the only ``nonlocal" violations of the CHSH inequality are counterfactual in nature. This should perhaps be expected being that measurements in QM are effectively ``classical" and therefore ``nonlocal factual" violations of the CHSH is nonsensical in these definitions.  

One point of interest is that the MOPA clashes with the usual motivation for the factorization condition $p(a,b|x,y,\lambda)=p(a|x,\lambda)p(b|y,\lambda)$ and its marginalization (\ref{margfactor}) that inevitably satisfy the CHSH inequality. Before a factorization condition could be stipulated in the MOPA one would need to first address, who is assigning the distribution and what information is this observer privy to at that time? That is, the factorization condition needs an observer label ($A$, $B$, or $C$) and perhaps the inclusion of the double vertical bar notation to divide factual and counterfactual propositions. The MOPA does not represent locality with a factorization condition, so factorization is not an obvious step in the MOPA. Because locality is represented as the factual account of observations made by the observers, this analysis coherently gels probability theory and SR.

The \emph{no-signaling} condition $p(a|x,y)\stackrel{n.s.}{=}p(a|x)$ in the MOPA may be stated more exactly.  If $y$ is known factually by Alice, $p_A(a||x,y)\stackrel{n.s.}{=}p_A(a||x)$, it implies $y$ was indeed signaled by Bob, which is then local. If instead $y$ is \emph{not signaled}, one could consider the value of $y$ counterfactually, and then $p_A(a|y|x)\stackrel{n.s.}{=}p_A(a||x)$ is a statement about measurement setting independence \emph{if} the measurement setting were set to $y$ -- which then again is a statement that is independent of whether $y$ is local or not. In either case, the \emph{no-signaling} condition is better stated as a \emph{marginal measurement setting independence}. An interesting note is that statements which are known to be true (such as $b\vee\tilde{b}$) are factually known to be true locally and nonlocally by observers per the definition of factuality; therefore, marginalized variables have a factual nature over their disjunction as eventually one of them in principle may be learned to be factually true, yet one may still consider a subset of $b\vee\tilde{b}$ counterfactually before measurement.

The mystery that entangled states present, is not \emph{no-signaling} or \emph{marginal measurement setting independence}, but rather is the fact that QM generates a joint probability distribution,
\begin{eqnarray}
P_{QM}(\pm_{a},\pm_{b}|\theta_a,\theta_b)=\frac{1}{4}+\frac{\delta_{\pm_a,\mp_b}-\delta_{\pm_a,\pm_b}}{4}\cos(\theta_a-\theta_b),\nonumber
\end{eqnarray}
over a pair of observables $\pm_{a}$ and $\pm_{b}$ that is conditioned on an unfactorisable $\cos(\theta_a-\theta_b)$ term over measurement settings. The explicit reason, or understanding, of \emph{why} this unfactorisable correlation term is present in QM is not addressed by a MOPA. What is addressed is that because a nonlocal violation of the CHSH can only ``happen" counterfactually, factual experimental verifications of nonlocal violations of the CHSH are nonsensical. Local violations of the CHSH may occur factually or counterfactually; however, in these instances the correlations between the detector settings and outcomes may be attributed to local interactions (or perhaps local entanglement mechanisms \cite{PL}). Because at best a MOPA's probability distributions conditional on nonlocal propositions require their dependence to be counterfactual, it implies that the probability distributions given by QM have \emph{counterfactual} dependences on the measurement settings, that is,
\begin{eqnarray}
P_{QM}(\pm_{a},\pm_{b}|\theta_a,\theta_b)\rightarrow P_{QM}(\pm_{a},\pm_{b}|\theta_a,\theta_b|...),
\end{eqnarray}
as it can never be verified otherwise \emph{by any observer}. This implies that $P_{QM}$'s, which are usually used in local small scale experiments and thus do not require observer labels (a local observer symmetry of sorts), in general require observer labels. Using QM in a general setting thereby requires observer labels to fully describe experiments with nonlocal propositions, which is in favor of a probabilistic and epistemic interpretations of QM.

The nonlocal counterfactual CHSH inequality is a statement of what is expected to occur (it is a sum of expectation values) rather than what will actually occur or is currently occurring -- which then removes it somewhat from an ``element of reality" to a quantity of epistemology. At $t_{\pm}$, Alice measures her particle and from it can infer what combination of $\pm_b$ and $\theta_b$ Bob is most likely to report to her at $t_c$. Given the realist view of physics adopted here, that $I_{A\cup B}$ be free of contradictions, Bob cannot report zero probability events to Alice as they both agree with one another's distributions initially at $t_0$ (\ref{t_0}). It is this requirement for realism that keeps Bob's measured values in check with Alice's perspective and vice-versa. The set of zero probability events are known from the outset at $t_0$, and the situation remains so through $t_c$ given the initial conditions $\psi_0$ are known precisely and remain so.

Because the operation of a Stern-Gerlach device is to entangle the spins of particles with their positions such that they may be detected on a screen, for a fully interpretable analysis of the Bell experiment in the MOPA, the positional part of the wavefunction $\ket{\Psi}$ should be taken into account during Bell experiments such that particle spins $\ket{\chi}\rightarrow\ket{\chi,\Psi}$, which allows arguments about location and locality to be made with more rigor. The extra structure provided by $\ket{\Psi}$ gives additional positional information that is neglected if positions are marginalized. Having positional information would allow one to better represent locality arguments in the mathematics rather than having to appeal to linguistics as is widely done and was done here. 

\begin{acknowledgements}
I benefited from numerous conversations throughout the writing of this article with Nick Carrara, Ariel Caticha, and Cai Waegell -- Thank you all. I would also like to thank the rest of our information physics research group at UAlbany.
\end{acknowledgements}

\end{document}